\documentstyle[aps]{revtex}
\newcommand{\wpin}{\relax}
\newcommand{\graphin}{\relax}
\newcommand{\tight}{\relax}
\def\[{\begin{equation}}
\def\]{\end{equation}}
\def\({\begin{equation}}
\def\){\end{equation}}
\def\mede#1{\langle\,#1\,\rangle}
\def\d{\hbox{d}}

\def\part{\partial}

\def\gsim{>\approx}
\begin{document}

\wpin

\title{%
Emergent Traffic Jams
}

\author{%
Kai Nagel${}^{a,b}$ and Maya Paczuski${}^a$\\
${}^a$ Brookhaven National Laboratory\\
${}^b$ Center for Parallel Computing ZPR, University~of~Cologne,
50923~K\"oln, Germany, email~kai@zpr.uni-koeln.de , %
maya@cmt1.phy.bnl.gov \\
\today
}

\maketitle

\begin{abstract}
We study a single-lane traffic model that is based on human driving
behavior.  The outflow from a traffic jam
self-organizes to a critical state of maximum throughput.  Small
perturbations of the outflow far downstream create emergent traffic
jams with a power law distribution $P(t) \sim t^{-3/2}$ of lifetimes,
$t$.  On varying the vehicle density in a closed system, this critical
state separates lamellar  and jammed regimes, and
exhibits $1/f$ noise in the power spectrum.  Using random walk
arguments, in conjunction with a cascade equation,
we develop a phenomenological theory that predicts the
critical exponents for this transition and explains the
self-organizing behavior.  These predictions are consistent with all
of our numerical results.
\end{abstract}

   \section{Introduction}
%
Traffic jams are annoying, and they have negative economic impact.
For example, it may be noted that in 1990 (1980), 14.8\% (16.4\%)
of the U.S. GNP was absorbed by passenger and
freight transportation costs~\cite{ Eno}.
Rather than increasing the supply of transportation, perhaps by adding
new highways or a train-based transit system,  or decreasing the
demand for transportation, for example by making it more expensive,
 it is desirable to use existing transportation
structures as efficiently as possible.  One would, perhaps,
want to keep a freeway in the regime of maximum vehicle
throughput.

However, it turns out that this regime is not very well understood.
Recent numerical simulations using grid based particle models for
traffic flow have found indications for a phase transition separating
low-density lamellar flow from high-density jammed behavior where
particles either stop moving or move very slowly~\cite{ Biham,
Takayasu.Takayasu, Nagel.92.Schrecki}.  It has been observed numerically
that this transition occurs at or near the point of maximum
throughput~\cite{ Nagel.94.ltimes} and that the flow behavior in this
region is complex.  Continuum fluid-dynamical approaches similarly predict
instabilities in this region~\cite{ Kuehne,
Kerner.Konh, Helbing}, consistent with real world
observations~\cite{ Hall.mea, Treiterer.hysteresis}.

Here we demonstrate that maximum throughput corresponds to a
percolative transition for the traffic jams.  It occurs at the point
where emergent traffic jams are barely able to survive indefinitely.
This implies that the
 intrinsic flow rate for vehicles leaving a jam equals maximum
throughput. As a result, the outflow from a large jam (at large
distances or times) self-organizes to the maximum throughput critical
point.  Numerical results show that slow perturbations in the outflow
lead to traffic jams, downstream, of all sizes -- a particularly
simple example of self-organized criticality (SOC)~\cite{ Bak.SOC}.
   If the system is ``driven'' with more
frequent random perturbations, then the jams will interact.  This induces
a finite correlation length for the jams and  pushes the system off
 criticality.
Similarly, the size of a jam induces a finite size cutoff in its
outflow.  These considerations imply that traffic in a complicated
network is likely to be poised near the critical state determined by
the largest jam in the system, and thus susceptible to small
perturbations.  The characteristic power law associated with the jam
lifetimes makes prediction of flow behavior more
difficult.
Steps that are taken to reduce random
fluctuations/perturbations, such as cruise control or automatic
car-following systems, in fact, push the traffic network closer to
its underlying critical point, thereby
making it {\it more} likely to have large
jams.

We study a simplified version of an original discrete model proposed
by Nagel and Schreckenberg~\cite{ Nagel.92.Schrecki}. This simplification
can be described as a ``cruise control limit'', since at sufficiently
low density all vehicles move deterministically at maximum allowed
velocity.  This deterministic motion is interrupted by small
perturbations at a vanishingly slow rate; i.e., the system is
allowed to relax back to a deterministic state before it is kicked
again.  The emergent
traffic jams are the transient response to the perturbation.

In the model, the forward motion of vehicles
in a single lane is mimicked by the forward motion of particles on a
one dimensional lattice.  The essential features of this model are:
 a) hard-core particle dynamics
 b) an asymmetry between acceleration and deceleration which, in
connection with a parallel update, leads to clumping behavior and jam
formation rather than smooth density fluctuations
 c) a wide separation between the time scale for creating small
perturbations in the system and the relaxational dynamics, or the
lifetime of the jams. The model is studied
with both closed and open boundary conditions.

This model exhibits behavior that is characteristic of
granular systems ~\cite{ Biham, granular,
Baxter.1.f.granular,
Ristow.1.f.granular, Peng.1.f.granular.1,
Peng.1.f.granular.2}.  These include phenomena ranging from the
rather mundane example of flow of sand in an hour glass
{}~\cite{ Schick.1.f.granular} to the
large scale structure of the universe~\cite{ universe}.
Recent studies of clustering instabilities in
one-dimensional  many particle systems in which
particles interact via inelastic collisions
{}~\cite{ inelastic} may also be related.

In Section II, the traffic model is defined, and its current-density
relation  is derived.
The outflow from a large jam is marked by a power law
scaling of the distribution of jam lifetimes with exponent $3/2$.
This outflow operates at the point
of maximum throughput.
Section III presents random walk arguments, which are exact for a version
of the model with maximum velocity, $v_{max}=1$.  This
theory predicts the critical exponents for the emergent jams.  The
number of jammed vehicles, $n$,
 scales with time as $t^{1/2}$.  The space-time jam size (or mass of the
jam cluster) $s \sim n t$, and the spatial extent $w  \sim  n$.  On varying
the density, $\rho$, away from the maximum throughput value, $\rho_c$,
the jams have a characteristic lifetime $t_{co}$, or cutoff, which
scales as $t_{co} \sim (\rho_c - \rho)^2$.  It is important to note
that jams with $v_{max} >1$ are allowed to branch, unlike $v_{max}=1$.
In Section IV, this branching behavior is analyzed in terms of a cascade
equation for the size distribution of intervals between parts of
the jam.  The distribution of interval sizes, $x$, is predicted to decay as
$1/x^2 $.  This result suggests that the jams are marginally dense
and the random walk theory is valid up to logarithmic corrections,
e.g. $w \sim t^{1/2} \ln t$.
  Also, since the jams drift backwards, this  distribution
of interval sizes gives rise to $1/f$ noise in
the power spectrum of local activity.
In Section V, we present the rest of our numerical results.
These  results are  consistent with our phenomenological
theory.  In Section VI, we discuss the
potential relevance of this work to real traffic.

   \section{The model}

The closed model is defined on a one-dimensional array of length $L$,
representing a single-lane freeway.  Each site of the array can be in
one of $v_{max}+2$ states: It may be empty, or it may be occupied by
one car having an integer velocity between zero and $v_{max}$.  This
integer number for the velocity is the number of sites each vehicle
advances during one iteration.  Movement is restricted to occur
``crash-free''.  Unless otherwise noted, we choose $v_{max}=5$, but
any value $v_{max} \ge 2$ gives the same large scale behavior
 when lengths are rescaled by a short distance
cutoff.  This short distance cutoff corresponds roughly to the typical
distance required for a vehicle starting at rest to accelerate to
maximum velocity.

For every  configuration of the model, one iteration
consists of the following steps, which are each performed
simultaneously for all vehicles (here, the
quantity $gap $ equals the number of empty sites
in front of a vehicle):\begin{itemize}

\item
A vehicle is  stationary when it travels at maximum velocity
$v_{max}$ and has free headway: $gap \ge v_{max}$.  Such a vehicle
just maintains  its velocity.

\item
If a vehicle is not stationary, it is jammed.
The following two rules are
applied to jammed vehicles:\begin{itemize}

\item
{\bf Acceleration of free vehicles:} With probability~1/2, a vehicle
with $gap \ge v+1$ accelerates to $v+1$, otherwise it keeps the
velocity~$v$.  A vehicle with $gap = v$ just maintains its velocity.

\item
{\bf Slowing down due to other cars:} Each vehicle with $gap \le v-1$
slows down to $gap$: $v \to gap$.  With probability~1/2, it overreacts
and slows down even further: $v \to \max[gap-1,0]$.

\end{itemize}

\item {\bf Movement:} Each vehicle advances $v$ sites.

\end{itemize}
Randomization takes care of two behavioral patterns:
(i)~Non-deterministic acceleration.  This is the source of the scaling
behavior of the jam lifetimes. (ii)~Over-reactions when slowing down.
This is
considered to be
 realistic with respect to real traffic~\cite{ Leutzbach, Wiedemann}.
For clarity, a formal version of the velocity update is given in the
appendix.

While in the original model studied by Nagel and Schreckenberg~\cite{
Nagel.92.Schrecki, Nagel.94.ltimes}, vehicles at $v_{max}$ slowed down
randomly with probability $p_{free}$, here only the jammed vehicles
move nondeterministically.  This corresponds to the $p_{free}
\rightarrow 0$ limit, or the ``cruise control limit'', of the previous
model and completely separates the time scales for
perturbing the system and the system's response.

Our fundamental diagram, or current-density relation, $j(\rho )$, was
determined numerically as shown in Fig.~\ref{fdiag} for a closed
system of size~$L = 30\,000$.  Starting with a random initial
condition with $N$~cars (i.e.\ $\rho=N/L$) and after discarding a
transient period of $5\cdot 10^5$~iterations, we measured
$\mede{j}_L(t) = \sum_{i=1}^N v_i / L$ every 2500 time-steps up to the
$3 \cdot 10^6$th iteration.  Each data point corresponds to the
average over current measurements for a single initial condition, with
the following exception: When a run becomes stationary (i.e.\ no more
jammed cars in the sense of the definition above), then the future
behavior is predictable.  In this case, the run is stopped, and the
current will be equal to $j_{det} = v_{max} \cdot \rho$, see below.

For a spatially infinite system, the following results hold: For $\rho
< \rho_c $, jams present in the initial configuration are eventually
sorted out and the stationary deterministic state is jam free with
every vehicle moving at maximum velocity.  Thus in the lamellar regime
the current is a linear function of density with slope $v_{max}=5$.
Lamellar behavior is observed up to a maximum current
$j_c(\rho_c)$. For $\rho > \rho_c$, and $\rho < \rho_{det,max}$
(defined below) the system is bistable.  Starting from an initial
configuration which has many jams, the jams in this case are never
sorted out.  The steady state is an
inhomogeneous mixture of jam free regions and higher density jammed
regions.  Clearly, these jammed regions decrease the average current
in the system.  It is possible, nevertheless, to prepare initial
configurations that have no jams.  Since all motion is deterministic
in this state, the steady state will also have no jams and the current
will still be a linear increasing function of $\rho$ (the dotted line
in Fig.~\ref{fdiag}).  This is possible up to densities of
\[
\rho_{det,max} = { 1 \over v_{max} + 1 } \ ,
\]
leading to a maximum current of
\[
j_{det,max} = { v_{max} \over v_{max} + 1 } \ .
\]
This clearly is much higher than the current $j_c$ for random initial
conditions.  It is in this sense that our system is bi-stable (cf.\
also~\cite{ Takayasu.Takayasu}).  This effect allows us to produce
outflows with densities above $\rho_c$.

Above $\rho_c$, the current-density relation can be derived by
assuming that the system phase separates into jammed regions separated
by jam free gaps.  The jam free gaps are the outflow of a jam and thus
have current $j_c(\rho_c)$, as argued in the next section.
Conservation of the number of cars and of volume
\cite{ explanation} leads to
\[
j = j_c - {(\rho - \rho_c)(a j_c - v_j) \over 1 - a \rho_c} \quad ,
\]
where $a$ is the average number of lattice sites per jammed vehicle,
and $v_j$ is the average velocity ($<v_{max}$) of a jammed vehicle
(see~\cite{ Bando.jpn.car.foll} for a similar calculation).  Thus, the
current-density relation is linear both above and below the critical
point, as demonstrated in Fig.~\ref{fdiag}.

The discontinuity in the current at the critical point, as seen in the
figure, is a finite size effect due to the fact that each point
in the figure represents a single initial configuration.
 In a finite system, there is a
finite probability that even a system with supercritical density $\rho
> \rho_c$ finds the deterministic state, and then has a current of
$j_{det} > j_c$.

\subsection{The outflow from a jam occurs at maximum throughput}

A striking feature of the model is that  maximum
throughput is selected automatically when the left boundary condition
is an infinitely large jam and the right boundary is open.  This
situation was  described for the original model
in~\cite{ Nagel.94.ltimes}.  An intuitive explanation is that maximum
throughput cannot be any higher than the intrinsic flow rate out
of a jam.  Otherwise the flow rate into a jam would be higher than the
flow rate out,  and the jam would be stable in the long time limit, thus
reducing the overall current.
By definition, of course, maximum throughput cannot be lower than
this intrinsic flow rate.

In Fig.~\ref{pixel-J5}, the cars on the left flow out from a region of
high density where they move with zero velocity.
This high density region is not plotted here; only the interface
or front separating the high density region and its deterministic
outflow is plotted.  This is the branched structure on
the left hand side of the figure.  The vehicles flowing
out of the large jam ultimately relax to the deterministic state
when they have move sufficiently far away from the jam.

This feature of maximum throughput selection is characteristic of
driven diffusive systems~\cite{ Carlson, Krug, Derrida.etc}.  However, in
our case the left boundary condition is unusual: the front of the infinite
jam drifts backward in time.   If the left boundary
is fixed in space and vehicles are inserted at velocities less than
$v_{max}$, then the outflow from a jam cannot reach maximum throughput
(cf.\ bottleneck situation in~\cite{ Nagel.92.Schrecki, Nagel.94.Herrmann}).
%
%
This point warrants further investigation, since it corresponds to the
real world observation that disturbances which are fixed in space,
such as bottlenecks or on-ramps, lead to much lower throughput
downstream than would be possible
theoretically~\cite{ Kanaan.lower.throughput.on.ramps}.

\subsection{Traffic jams in the outflow show self-organized criticality}

The outflow situation, as described above, produces deterministic flow
asymptotically at large distances.  This means that sufficiently far
downstream from the large jam, the jam flow has sorted itself out into
deterministic flow.  In the deterministic region, one car is randomly
perturbed by reducing its velocity to zero.  Many different choices
for the local perturbation, however, give rise to the same large scale
behavior.  The perturbed car eventually re-accelerates to maximum
velocity.  In the meantime, though, a following car may have come too
close to the disturbed car and has to slow down.  This initiates a
chain reaction -- the emergent traffic jam.

Fig.~\ref{pixel-J5} also shows the first 1400~time steps of such an
emergent jam, as the structure on the right hand side of the figure.
 Qualitatively, the jam
clearly shows a tendency to branch with complex internal structure and
a fractal appearance ~\cite{ Mandelbrot}.
The emergent traffic jams drift backwards; 
so
it is possible for a sufficiently long lived emergent jam to
eventually intersect with the outflow jam interface, on the
left in Fig.~\ref{pixel-J5},  that is itself
becoming broader with time.  It is likely that the branching behavior
of the emergent jams is the same as the branching behavior of the
original jam interface.  In this work, however, we do not explicitly
study the interface.  Contrary to the figure, in the computer code,
the interface region to the left and the
emergent jam to the right are kept
completely separate using methods described in Appendix I.

A jam is sorted out when the number of jammed cars is zero.  This
defines the lifetime, $t$, of an emergent traffic jam.  In order to
obtain statistics for the properties of noninteracting traffic jams,
the deterministic outflow is  disturbed
again, after the previous jam has died out.
In our simulations we measure the lifetime distribution, $P(t)$,
the spatial extent $w$ of the
jam, the number of jammed vehicles $n$,
 and the overall space-time size $s$ (mass) of the jam.
These properties of the traffic jam are analogous to
other branching processes such as directed percolation~\cite{
Grassberger.79.Torre}, branching annihilating random walks~\cite{
Jensen.94.BAW}, or  nonequilibrium lattice models~\cite{
Jensen.93.Dickman}, although the precise behaviors are different.

Fig.~\ref{pixel-J4} shows 1400~time steps in the middle of the life of
a larger jam. Here, vehicles that are stationary are no longer
shown; the plot  only shows the ``particles'', or jammed
vehicles, that propagate the disturbance.

For a quantitative treatment, we start by measuring the probability
distribution of jams as a function of their lifetime, $t$.
Fig.~\ref{soc-ltimes} shows that for $t \gsim 100$ this
distribution follows a power law
\[
P(t) \sim t^{-(\delta+1)}
\qquad\hbox{ with }\qquad (\delta+1) = 1.5 \pm 0.01 \ ,
\]
very close to $\delta = 1/2$.
This figure represents averaged results
of more than 60\,000~jams.

Here scaling is observed over almost four orders of magnitude as
determined by our {\it numerically imposed} cutoff: For this figure,
if jams survive longer than $10^6$ time steps, they are removed from
the data base.
It is very important to note that these emergent jams
are precisely critical.  Their power law scaling persists up to any
arbitrarily large, numerically imposed cutoff.  The lifetime
distribution is related to the survival probability~$P_{surv}(t)$ by
\[
P_{surv}(t) = \int_t^\infty \d t' P(t') \sim t^{-\delta}
\qquad\hbox{ for }\qquad \delta > 0 \ .
\]
We again  emphasize that no external tuning is necessary to observe
this scaling behavior.  The outflow  from the
infinite jam self-organizes to
the critical state.

   \section{Random Walk Arguments for Critical Behavior}

It is, perhaps, surprising that such a seemingly complicated structure as
shown in Fig.~\ref{pixel-J5} is described by such a simple apparent
exponent.  Numerically, the exponent $\delta + 1$ is conspicuously close to
$3/2$, the
first return time exponent for a one-dimensional random walk.  In fact,
for $v_{max}=1$ this random walk picture is exact, as shown below.

Let us consider a single jam in a large system with $v_{max}=1$.  The
vehicles in the jam form a queue, and all of these cars have velocity
zero.  When the vehicle at the front of the jam accelerates to
velocity one, it leaves the jam forever.  The rate at which vehicles
leave the jam is determined by the probabilistic rule for
acceleration.  Vehicles, of course, can be added to the jam at the
back end.  These vehicles come in at a rate which depends on the
density and velocity of cars behind the jam.  Given the rules for
deceleration, the spacing between the jammed cars is zero so the
number of cars in the jam, $n$, is equal to the spatial extent of the
jam, $w$.  This contrasts with the branching behavior for $v_{max}
>1$.  The probability distribution, $P(n,t)$, for the number of cars
in the jam $n$ at time $t$ is determined by the following equation:
\[ P(n,t+1) = (1 - r_{in} - r_{out})P(n,t) + r_{in}P(n-1,t)
+ r_{out}P(n+1,t) \quad .\label{rate}
\]
Here, the quantities $r_{in}$ and $r_{out}$ are phenomenological parameters
that depend on the density behind the jam and the rate at which cars leave
a jam.  They are independent of  the number of cars in the jam.
For large $n$ and $t$, one can take the continuum limit of Eqn.~\ref{rate}
and expand to lowest order
\[ {\partial P \over \partial t} = (r_{out} - r_{in}){\partial P
\over \partial n} + {r_{out} +r_{in} \over 2}{\partial^2 P \over
\partial n^2} \quad .\label{continuum}
\]

When the density behind the jam is such that the rate of cars entering
the jam is equal to the intrinsic rate that cars leave the jam, then
the first term on the right hand side vanishes, and
the jam queue is formally equivalent to an
unbiased random walk in one dimension~\cite{ Feller}, or the diffusion
equation.
The first return time of the walk then corresponds to the
lifetime of a jam.  This leads immediately to the result $P(t) \sim
t^{-3/2}$ for the lifetime distribution.

This argument shows that the outflow from an infinite jam is in
fact self-organized critical.  This can be seen  by noting
that the outflow from a large jam occurs at the same rate as the
outflow from an emergent jam created by a perturbation.
This also shows that maximum throughput corresponds to
the percolative transition for the traffic jams.  Starting from random
initial conditions in a closed system, the current at long times is
determined by the outflow of the longest-lived jam in the system.

When $r_{in}=r_{out}$, one also finds from
Eqn.~\ref{continuum} that $n \sim t^{1/2}$ and the size of
the jam $s \sim n t \sim t^{3/2}$.  If the density in the
deterministic state is below the critical density $\rho_c$, then the
jams will have a characteristic lifetime, $t_{co}$, size $s_{co}$,
number $n_{co}$, etc.  From Eqn.~\ref{continuum}, $t_{co}
\sim n_{co} (r_{out} - r_{in})^{-1}$.  Assuming that near the critical
point $r_{out} - r_{in} \sim \rho_c - \rho$, then using $n_{co} \sim
t_{co}^{1/2}$ leads to
\[ t_{co} \sim (\rho_c - \rho)^{-2}
\quad .
\]
If the left boundary condition is such that
$\rho > \rho_c$, vehicles on average enter the emergent
jam at a faster rate than
they leave.  In this case, there is a finite probability
to have an infinite jam, $P_{\infty}$,
which vanishes as $\rho \rightarrow \rho_c$
as
\[ P_{\infty} \sim (\rho - \rho_c)^{\beta} \quad .
\]
In a closed system,
 the steady-state density of jammed cars, $\rho_j = \rho - \rho_c$,
so that
the order parameter exponent is trivially $\beta =1$.  From the random walk
Eqn. (~\ref{continuum}), and in analogy with other branching processes
such as directed percolation~\cite{ Grassberger.79.Torre}, $P_{surv}$
follows a scaling form
\[
P_{surv}(t,\Delta) \sim t^{-\delta} \, f(t \, \Delta^{\nu_t}) \ ,
\] near the
critical point.  Here $\Delta \equiv |\rho - \rho_c|$ and $t_{co} \sim
\Delta^{-\nu_t}$.
 From this scaling relation,
${\beta} = \delta \nu_t $.  For $v_{max} = 1$, $\delta = 1/2$, $\nu_t = 2$, and
again
$\beta =1$.

The number of jammed vehicles, ${\bar n}$,
averaged over all  jams, including those that die out, has the scaling form
\[{\bar n} \sim t^{\eta}g(t \, \Delta^{\nu_t}) \ .
\]
The number of jammed vehicles averaged over surviving jams,
scales with a different exponent
\[n(t)= {\bar n}(t)/P(t)
\sim t^{\eta + \delta} \ .
\]
The mapping to the random walk gives $\eta = 0$.
The cluster width, averaged over surviving clusters, scales as
$w \sim t^{1/z}$, and the mapping to the random walk gives $z=2$.  The
average cluster size $ s \sim t^{\eta + \delta + 1}$; $s
\sim t^{3/2}$ in the random walk case.

In the numerical measurements, we averaged the quantities $t = $
lifetime of the cluster, $w =$ maximum width of cluster during cluster
life, $n =$ maximum number of simultaneously jammed vehicles during
cluster life, $s =$ total number of jammed vehicles during cluster
life.

Our theoretical results should describe the emergent traffic jams not
only at $v_{max} =1$ but also for any $v_{max} >1$ as long as the
traffic jam itself remains dense.  If this is the case, then the
dynamical evolution is determined solely by the balance of incoming
and outgoing vehicles as described by Eqn.~\ref{continuum}.  The
ratio~$ w / n$ should go to a finite constant at large
times if the theory is valid.  If the emergent jams break up into a
fractal structure, and $ w / n$ diverges, internal
dynamics must also be included.  Since the jams displayed in
Figs.~\ref{pixel-J5} and~\ref{pixel-J4} appears branched and at least
qualitatively fractal, one might doubt that such a simple theory could
describe this behavior.  Nevertheless, the close numerical agreement
of the lifetime distribution exponent for the SOC behavior suggests
the possibility that the random walk theory is a valid description of
the branching jam waves.

\section{A Cascade Equation for the Branching Jams}

We now analyze the branching behavior of jams with $v_{max} >1$ in
terms of a phenomological cascade equation.  A very large emergent
jam, at a fixed point in time, consists of small dense regions of
jammed cars, which we call subjams, separated by intervals, ``holes'',
where all cars move at maximum velocity.  If the jam is dense, then
the holes have a finite average size.  Otherwise, the jammed vehicles
may comprise a fractal with dimension $d_f < 1$.  We will consider the
subjams to have size one.

Holes between the subjams
 are created at small scales by the probabilistic rules
for acceleration.  Each subjam can create small holes in front of it.
We will ignore the details of the injection mechanism, and assume that
there is a steady rate at which small holes are created in the interior
of a very long lived jam.  We also assume that the
 interior region of a long-lived jam reaches a steady state
distribution of hole sizes.
We do not explicitly study the distribution
of hole sizes at small scales.

In order to determine the asymptotic scaling of the large holes
in the interior of a long-lived jam,
it is necessary to isolate the dominant mechanism in the cascade process for
large hole generation.  This mechanism is
the dissolution of one subjam.  When one subjam
dissolves because the cars in it accelerate to maximum velocity,
the two holes on either side of it merge to form one larger hole.  Holes at
any large scale are created and destroyed by this same process.
This mechanism links different large scales together, and we propose
that it gives the leading order contribution at large hole sizes.
In the steady state, the creation and destruction of large holes must
balance.  This leads to a cascade equation for holes of size $x$:
\(
\sum_{u=x+1}^{\infty}<h(x)h(u-x)> = \sum_{x'=1}^{x-2}<h(x')h(x-x'-1)>
\quad . \label{cascade}
\)
Here, the angular brackets denote an ensemble average over all holes in the
jam, and the quantity $h(x)h(u-x)$ denotes a configuration where
a hole of size $x$ is adjacent to a hole of size $(u-x)$.  The right hand side
of this equation represents the rate at which holes of size $x$ are
created, and the left hand side represents the rate at which holes of
size $x$ are destroyed.

Now, we make an additional ansatz; namely, for
large $x$, $<h(x')h(x-x'-1)>= G(x)$, independent of $x'$ to leading order.
That is, to leading order the probability to have two adjacent holes,
whose sizes sum to $x$ is independent of the size of either hole.
$G(x)$ then also scales the same as $P_h(x)$, the probability to have a hole
of size $x$.
Thus Eqn.~\ref{cascade}, to leading order, can be written
\(
\sum_{u=x}^{\infty} G(u) \sim xG(x)   \quad.
\)
Differentiating leads to
\(
x{\partial G(x) \over \partial x} = -2G(x) \quad ; \quad G(x) \sim {1\over x^2}
\quad .
\)
Thus the distribution of hole sizes decays as
\(
P_h(x) \sim x^{-\tau_h} \quad ; \quad {\rm with \ \ } \tau_h =2 \quad .
\)
It is interesting to note that the cascade equation (\ref{cascade}) is
identical to the dominant mechanism in the exact cascade equation for
forests in the one-dimensional forest fire model
{}~\cite{ Paczuski.93.Bak}.
The exponent
$\tau_h =2$ is the same as the distribution
exponent for the  forests, which has been obtained exactly
{}~\cite{ Drossel.1dff}.
 Curiously, $\tau_h = 2$  can also
be regarded as another example of Zipf's law ~\cite{ Zipf}.

The exponent $\tau_h$ is related to the fractal dimension~$d_f$
of jammed vehicles by
\[
\tau_h = 1 + d_f \ ,
\]
as long as $\tau_h \le 2$~\cite{ Maslov.95.P.B}.
 Thus, $\tau_h < 2$ implies that
the equal time cut of the jam clusters is fractal, otherwise not.  The
point $\tau_h =2$ is the boundary between fractal and dense behavior.
At this special point, the random walk theory
can still be expected to apply,
although with logarithmic corrections.

The width of an emergent jam, at a given point in time,
 $w(t)$, can be expressed as
\[
w(t) = { n(t) \over w_j}\left(w_j + \int \d x \, x \, P_h(x,t)\right) \ .
\]
Here, $w_j$ is the average width of a subjam; it is ${\cal O}(1)$.
The quantity $P_h(x,t)$ is the probability distribution
to have a hole of size $x$ in
a jam that has survived to time $t$.  It is natural to assume that this
distribution corresponds to  $P_h(x)$ up to a cutoff which grows with $t$.
Inserting the expression for $P_h(x)$ gives
\[
w(t) \sim  n(t) \, \left( 1
+ \int_1^{x^*} \d x \, x^{1-\tau_h} \right) \ ,
\]
where the upper bound $x^*$
represents a time-dependent cutoff.
Using $\tau_h=2$,
$n \sim t^{\delta + \eta}$,
and assuming $x^* \sim t^c$
gives
\[
w(t) \sim t^{\delta+\eta} \, ( 1 + c \, \log t )
\qquad\hbox{ for } \tau_h = 2
\quad .
\]
In other words, if $\tau_h = 2$, as the above arguments suggest,
spatial quantities such as
$w(t)$ will exhibit logarithmic corrections to the random walk
results.   In the following section, we test these theoretical predictions
with further numerical studies.

   \section{Simulation results }

 We now present the rest of our numerical results.
Unless otherwise noted, these results were obtained for systems with
$v_{max} = 5$.

\subsection{At The Self-Organized Critical Point}

We study the critical properties of the outflow of a large jam by
driving it with slow random perturbations as described in Sec.~II.
Numerically, we  find (Fig.~\ref{n-t})
\(
n(t) \equiv \mede{n}_{surv}(t) \sim t^{\eta + \delta}
\qquad \eta + \delta  =  0.5 \pm 0.1
\)
and (Fig.~\ref{s-T})
\(
s(t) \sim  n(t) t \sim t^{1 + \eta + \delta}
\qquad 1 + \eta + \delta = 1.5 \pm 0.1
\)
in agreement with the random walk predictions.  However, the
simulations do not converge to power law scaling before $t
\simeq 3 \cdot 10^4$, and since the simulation is cut off at $t = 10^6$,
the exponents are obtained from less than two orders of magnitude
in~$t$.  Figs.~\ref{n-t} and~\ref{s-T} contain the averaged results of
more than 160\,000~avalanches, typically corresponding to
approximately 200~workstation hours (see Appendix and figure captions
for further information).

\subsection{Off Criticality}

By changing the left boundary condition (i.e.\ the inflow condition)
of the open system, simulations were performed both above and below
the critical point.  This is achieved by replacing the mega jam by the
following mechanism: Vehicles are inserted with $v=v_{max}$,
 at a fixed left boundary. After each vehicle,
$v_{max}$ sites are left empty and then the
following sites are attempted to be occupied with probability
$p_{insert}$ until a site is occupied.  The rate $p_{insert}$
determines an average density $\rho$ by
\[
\rho = { 1 \over v_{max} + {1 / p_{insert}} } \ ,
\]
which can go as high as $\rho = \rho_{det,max} = 1/6 = 0.16666\ldots$
for $v_{max}=5$, much higher than the critical density of $\rho_c \approx
0.0655$.

We have measured the survival probability, $P_{surv}(t)$ on varying
the density as shown in Fig. 7.  Based on the same data,
we have performed data collapse for the lifetime distribution
$P(t)$ on varying the density, as shown in Fig.~\ref{collaps}.
By plotting $P/t^{-(\delta + 1)}$ vs.\ $t \, \Delta^{\nu_t}$ with the
exponents $\delta+1 =1.5$, $\nu_t =2$ was determined by the qualitatively
best collapse.  The accuracy of this method is not very high, though,
so that the conclusion from the numerical results is no better than
\(
\nu_t = 2 \pm 0.2  \ ,
\)
which, again, in agrees with our random walk predictions.

\subsection{Explaining Previous Results}

These findings put us in a position to view simulation results of the
original model~\cite{ Nagel.94.ltimes} in a
%
new context (see
also~\cite{ Nagel.94.Rasmussen}).  In that model, multiple jams exist
simultaneously.  Jams start spontaneously and independently of other
jams because vehicles fluctuate even at maximum speed, as determined
by the parameter $p_{free} \neq 0$.

The original model displayed a scaling regime near the
(self-organizing) density of maximum throughput $\rho(j_{max})$, but
with an upper cutoff at $t \simeq 10^4$ which was observed to depend
on $p_{free}$.  We can now attribute this cutoff to the
non-separation of the time scales between disturbances and the
emergent traffic jams.  As soon as $p_{free}$ is different from zero,
the spontaneous initiation of a new jam can terminate another one.
Obviously, this happens more often when $p_{free}$ is high, which
explains why the scaling region gets longer when one reduces
$p_{free}$.  Dimensional arguments suggest that the cutoff in the
space-time volume, $V \sim w t$, should scale as $V_{co}p_{free}\sim
1$ (for $p_{free}\ll1$) since this implies that a new jam is initiated in a
space-time volume occupied by a previously initiated jam.  According
to the random walk picture $V \sim s$, so that $s_{co}\sim
p_{free}^{-1}$ and $t_{co} \sim p_{free}^{-2/3}$.  Measuring these
correlation lengths, however, is outside of the scope of the present
study.

   \subsection{Spatial Behavior}

So far, we have only shown simulation results for exponents describing
the evolution of the number of vehicles, but not their distribution in
space.  Here, our simulation results are less conclusive.
The width $w(t)$ vs.\
$t$~(Fig.~\ref{w-t}) is, besides the convergence problems already
described, best approximated by an exponent
\(
{1 \over z} = 0.58 \pm 0.04
\)
instead of $1/2$.  Measurements of other relations (e.g.\ $w$ vs.\
$n$; not shown) confirm these discrepancies for the spatial behavior
for branching jam clusters with $v_{max} >1$.  However, the form $w(t)
\sim t^{1/2} \ln t$ vs.\ $t$~(Fig.~\ref{w-t}) is also consistent with
the numerics.


In an effort to resolve this question, we analyzed large jam
configurations.  We ran simulations with $v_{max}=2$ until a cluster
reached a width of, say, $2^{13} = 8192$, and stored the configuration
of this time-step.  About 60 configurations of the same size were
used.  Measuring the distribution of holes inside the
configurations is consistent with the results from the cascade
equation presented earlier.

Fig.~\ref{N-holes} shows a plot of the probability distribution for
hole sizes, $P_h(x)$ vs.\ $x$, obtained from these configurations.  We
find
\(
P_h(x) \sim x^{-\tau_h} \qquad \tau_h = 1.96 \pm 0.1 \ ,
\)
which is indeed consistent with the prediction $\tau_h=2$ from the
cascade equation.

Nevertheless, our numerical results are not precise enough to distinguish
$\tau_h = 2$ from $\tau_h < 2$.  Nor do our measurements for the
width distinguish the  power
law fit with exponent $0.58$ from the theoretically plausible fit with
exponent $1/2$ and a logarithmic correction.

 \subsection{$1/f$ noise}

We measured the power spectrum by first recording the time series for
the number of  vehicles, $N_l(t)$, in a small segment of
length~$l$ in a closed system, and then taking the square of the
Fourier transform:
\[
S(f) = |N_l(f)|^2 = |FT[N_l(t)]|^2 \ .
\]
Since the jams have a finite drift velocity, the distribution of hole
sizes in space is translated into the same distribution of time
intervals for the activity.  In particular, the hole size distribution
in space translates to the first return time for jammed vehicles when
sitting at a fixed position in space.  It has been shown~\cite{ Maslov.95.P.B}
that given a distribution of first return times of activity
$P_{first}(t) \sim t^{-\tau_{FIRST}}$, the power spectrum scales as
$S(f) \sim 1/f^{\tau_{FIRST}-1}$.  Using the result
$\tau_h=\tau_{FIRST}=2$ this gives precisely a $1/f$ power spectrum
for the noise.  The power spectrum for the original model with the
parameter $p_{free} = 0.5$, $0.005$, $0.00005$ was measured in a
closed system near the critical density.  As shown in
Fig.~\ref{spectrum} the numerical results are in general agreement
with this prediction.  This result agrees  qualitatively with power
spectrum results for granular flow both in
experiments~\cite{ Schick.1.f.granular, Baxter.1.f.granular} and in
simulations~\cite{ Ristow.1.f.granular, Peng.1.f.granular.1,
Peng.1.f.granular.2}, and offers an alternative
explanation for
$1/f$-noise observed in traffic flow~\cite{ Musha.1, Musha.2}.

\section{Applications to real traffic}

With respect to real world traffic, much of this discussion appears
rather abstract.  A configuration of size $2^{13} = 8192$, as analyzed
in this work, corresponds to more than 100~km of undisturbed roadway,
a situation that rarely occurs in reality.  However, the
following  results
should be general enough to be important for traffic:
\begin{itemize}

\item
The concept of critical
phase transitions is helpful for characterizing real traffic behavior.
Open systems will tend to go close to a critical state that is determined
by the outflow from large jams.  This underlying
self-organized critical state corresponds to a percolative
transition for the jams; i.e. spontaneous small fluctuations can lead
to large emergent traffic jams.

\item
Interestingly, planned or already installed technological advancements
such as cruise-control or radar-based driving support will tend to
reduce the fluctuations at
maximum speed similar to our limit, thus increasing the range of
validity of our results.  One unintended consequence of these flow control
technologies is that, if they work, they will in fact push the traffic
system closer to its underlying critical point; thereby making prediction,
planning, and control more difficult.

\item
The fact that traffic jams are close to the border of fractal behavior
means that, from a single ``snapshot'' of a traffic system, one will
not be able to judge which traffic jams come from the same `reason'.
Concepts like queues~\cite{ Simao.queueing.models} or single waves do not make
sense when traffic is close to criticality. `Phantom' traffic jams
emerge spontaneously from the dynamics of branching jam waves.

\item
The fact that holes scale with an exponent around $-2$ means that, at
criticality, the jammed cars are close to not carrying any measure at
all.  The regime near maximum throughput thus corresponds to large
``holes'' operating practically at $\rho_c$ and $j_{max}$, plus a
network of branched jam-clusters, which do not change $\rho$ and $j$
very much.  The fluctuations found in the 5-minute-measurements of
traffic at capacity~\cite{ Hall.mea} therefore reflect the fact that
traffic flow is inhomogeneous with essentially two states (jammed and
maximum throughput).  The result of each 5-minute-measurement depends
on how many jam-branches are measured during this period.

\end{itemize}

    \section*{Appendix I Computational strategies and problems}

Computationally, we use a ``vehicle-oriented'' technique for most of
the results presented here.  Vehicles are maintained in an ordered
list, and each vehicle has an integer position and an integer velocity.
Since we model single lane traffic, passing is impossible, and the
list always remains ordered.


We simulate a system which is, for all practical purposes, infinite in
space.  The idea is comparable to a Leath algorithm in
percolation~\cite{ Leath}, which also only remembers the active part
of the cluster.

As we described earlier, a jam-cluster is surrounded by deterministic
traffic.  Let us assume that the leftmost car of this jam has the
number~$i_{left}$ and is at position $x_{left}$ (similar for the
rightmost car).  Cars are numbered from left to right; traffic is
flowing from left to right.

To the right of car~$i_{right}$, everything is deterministic and at
maximum speed, and, in consequence, nothing happens which can
influence the jam.  Therefore, we do not change the properties of the
jam if we do not simulate these cars.  Moreover, as soon as
car~$i_{right}$ becomes deterministic, it can never re-enter the
non-deterministic regime.  Therefore, car number $i_{right} - 1$
becomes the new rightmost car, and car number $i_{right}$ is no longer
considered for the simulation.

To the left of car~$i_{left}$, the situation is similar.  The only
information that we need is the sequence of the gaps $(gap_i)_i$ of
the incoming cars.  Just before car~$i_{left} - 1$ enters the jam, we
add one more car to the left, with $gap_{i_{left}-2}$.

It is obvious that, with this computational technique, the only
restriction for the spatial size is given by the memory of the
computer.  Since our model is one-dimensional, this has never been a
problem.

A remaining question is how to obtain the sequence of gaps $(gap_i)_i$
of the incoming cars, especially for the outflow situation.  One
possibility would be to first run another simulation of the outflow
from a mega-jam.  Cars leave this mega-jam, drive through a regime of
decreasing density, and eventually relax to the deterministic state.
One then records the gaps between these cars, writes them to a file,
and reads this file during the other simulation.  Apart from
technicalities (avoiding the intermediate file), this is the technique
we adopted in our simulations.

Our program runs with approximately $270\,000$~vehicle updates per
second on a SUN Sparc10 workstation; and since the critical density is
$\rho_c \approx 0.0655$, for $v_{max}=5$ this corresponds to $270\,000
/ 0.0655 = 4.1 \cdot 10^6$~site updates per second.  This includes all
time for measurements and for the production of the gaps of the
incoming cars.

We showed that our numerical results can not resolve the question
between logarithmic corrections for the width~$w(t)$ or an exponent
different from $1/2$, in spite of data obtained over six orders of
magnitude in time.  The reason for this is a large ``bump'' in the
measurements of the width.  Simulations of larger systems would have
been helpful.  The time complexity for our questions is $O(t)$: As
shown above, when averaging over all {\em started\/} clusters, the
number of active sites, $\mede{n}_{started}$, is constant in time:
$\mede{n}_{started}(t) \sim t^0$.  When $t_{co}$ is the numerically
imposed cut-off, then we perform for each started cluster in the
average $\alpha \, t_{co}$ updates of a vehicle.  In our experience,
$\alpha \approx 5$ for $v_{max}=5$.

In other words, in order to add another order of magnitude in time,
with the same statistical quality as before, we would need a factor of
10 more computational power.  However, each of our graphs already
stems from runs using 4~or more Sparc10 workstations for 10~days or
more.  And using a parallel supercomputer seems difficult: Standard
geometric parallelization is ineffective because most of the time the
jam-clusters are quite small, and in consequence all the CPUs
responsible for cars further away ``from the middle of the jam'' would
be idle.  More sophisticated load-balancing methods might be a
solution.

\section*{Acknowledgements}

This work was supported by the U.S. Department of Energy Division of
Materials Science, under contract DE-AC02-76CH00016. MP thanks the
U.S. Department of Energy Distinguished Postdoctoral Research Program,
KN the ``Graduiertenkolleg Scientific Computing K\"oln/St.~Augustin''
for financial support.  We thank Sergei Maslov, Per Bak, Sergei
Esipov, Chris Caplice, and Oh-Kyoung Kwon for useful discussions.  KN
thanks Brookhaven National Laboratory and Los Alamos National
Laboratory for their hospitality.  The ``Zentrum f\"ur Paralleles
Rechnen'' ZPR K\"oln provided most of the computer time.


\tight

\def\fig#1\par{\begin{figure}\caption{#1}\end{figure}}



\fig\label{fdiag}%
The fundamental diagram, $j(\rho)$, for $v_{max}=5$.  The dotted line
is valid for deterministic traffic, i.e.\ when the initial state is
prepared such that for each car $gap>v_{max}$ and $v=v_{max}$.  The
points are measurement results starting from random initial
conditions; each point corresponds to one run of a closed system of
length~$L=30\,000$ and an average over $2.5 \cdot 10^6$ iterations.

\fig\label{pixel-J5}
Outflow from a dense region (left); only the front, or interface,
from the dense region is shown as the structure on
the left hand side.  Dots represent vehicles which move to the
right.
The horizontal direction is space and the vertical direction (down)
is (increasing) time.   In the outflow region, an
emergent jam is triggered by a small disturbance.
This is the structure on the right hand side.
``Deterministic'' vehicles to the right of the emergent jam are
not plotted.

\fig\label{pixel-J4}%
  Space-time plot of an emergent jam.  The horizontal direction
is space and the verticle direction is time, as in Fig. (2).
 Only vehicles with $v < v_{max}$, i.e.\
  ``particles'',  are plotted.

\fig\label{soc-ltimes}%
Lifetime distribution $P(t)$ for emergent jams in the outflow region;
average over more than 65\,000 clusters (avalanches).  The dotted
line has slope $3/2$.  Numerically imposed cutoff at $t=10^6$.

\fig\label{n-t}%
 Number of jammed particles at time $t$, $n(t)$,
averaged over surviving clusters, in the outflow situation.
Numerically imposed cutoff at $t=10^6$; more than 165\,000 clusters
were simulated.  The dotted line has slope~$1/2$.

\fig\label{s-T}%
Mass of jam in space-time, $s(t)$, in the outflow situation, for the
same clusters as in Fig.~\protect\ref{n-t}.  Jams of similar lifetime
$t$ were averaged.  The dotted line has slope~$3/2$.

\fig\label{feed-ltimes}%
Survival probability for the jam clusters, $P_{surv}(t)$, for different
inflows.  Note that this distribution is highly sensitive to the
inflow, reconfirming that the outflow is indeed precisely critical.

\fig\label{collaps}%
Data collapse for the lifetime distribution of jams for the same data as
for Fig.~\protect\ref{feed-ltimes} with $\delta +1 = 1.5$ and $\nu_t = 2$.

\fig\label{w-t}%
Averaged maximum width of clusters, $w$, as a function of their
life-time, $t$.  The dotted line has slope $0.58$; the solid line
is a logarithmic fit $A \cdot t^{1/2} \cdot \log(t)$ where $A$ is a
free parameter.

\fig\label{N-holes}%
Probability distribution $P_h$ for hole-sizes $x$.  The dotted
line has slope $-2$.  The average is over 60~configurations, which all
have width $w=2^{13}$.  Contrary to all other figures in this
chapter, these results were obtained with $v_{max}=2$.

\fig\label{spectrum}%
Power spectrum, $S(f)$, smoothed by averaging, for a closed system of
length~$L=10^5$ and with $p_{free}=0.00005$.  Dotted line has slope
$-1$.

\graphin


\begin{thebibliography}{99}
\bibitem{ explanation}
For clarity, $N_f + N_j = N$, $L_f + L_j = L$, $N_j/L_j = 1/a$,
$N_f/L_f= \rho_c$, where $N_f$,$L_f$, and $\rho_c$ are
the number, length, and density, respectively, in the free flowing
phase, and $N_j$, $L_j$, and $1/a$ are the corresponding quantities
in the jammed phase.
\end{thebibliography}
\end{document}